\documentstyle[aps,prl,multicol,epsf]{revtex}
\begin{document}
\def\min#1{\smash{\mathop{\; \it min \;}\limits_{#1}}} 
\def\inf#1{\smash{\mathop{\; \it inf \;}\limits_{#1}}}
\def\sup#1{\smash{\mathop{\; \it sup \;}\limits_{#1}}}
\draft
\title{ A Note on Quantum Cloning in $d$ dimensions}
\author{ Paolo Zanardi
 \footnote{zanardi@isiosf.isi.it}} 
\address{ Institute for Scientific Interchange Foundation, Villa Gualino\\
Viale Settimio Severo 65, I-10133 Torino, Italy\\
and Unit\`a INFM, Politecnico di Torino,\\
Corso Duca degli Abruzzi 24, I-10129 Torino, Italy
}
\maketitle
\begin{abstract}
{The quantum state space $\cal S$ over a $d$-dimensional Hilbert space
is represented as a convex subset  of a $D-1$-dimensional sphere 
$S_{D-1}\subset {\bf{R}}^D$, 
where $D=d^2-1.$
Quantum tranformations ($CP$-maps) are then associated with the 
affine transformations of ${\bf{R}}^D,$ and $N\mapsto M$ {\it cloners}
induce polynomial mappings. 
In this geometrical setting it is shown that an optimal
 cloner can be chosen covariant and induces a map between reduced 
density matrices
given by a simple contraction of the associated $D$-dimensional Bloch
vectors. 
}
\end{abstract}
\begin{multicols}{2}
\narrowtext
\section{introduction}
The quantum {\it no-cloning} theorem \cite{WOZU} represents the most basic
difference between quantum and classical information theory.
It stems simply from the unitary character of any allowable evolution for a closed
quantum system. Since {\it perfect} copying of quantum information is forbidden
it is a relevant (conceptually as well as practically) question to ask
 how close one can get to that ideal (unphysical) process, and in what way.
More formally one has to face a complex optimization problem 
involving all allowed  quantum transformations between multipartite Hilbert spaces
($CP$ maps,\cite{KRA}).\\
Several papers, addressing this issue, have appeared recently.
Optimal fidelities and explicit forms for the cloning transformations
have been found \cite{BUHI},\cite{BUHI1},\cite{BUBR},\cite{GIMA}
 \cite{BRDI}, and connections with the Quantum State Estimation
problem has been made \cite{BREK}.
 These works are mainly focused on {\it qubit} (i.e. bi-dimensional)
systems (notably with the 
exception of reference \cite{WER} from which this note was inspired).
In this paper a few simple results are reported about the cloning problem
for an arbitrary $d$-dimensional quantum system, mostly obtained
in a geometric framework  (Generalized Bloch representation).
Although no explicit computations of cloning machines 
or cloning fidelities \cite{WER} appear,  we believe that the approach 
presented here 
deserves attention, in that it  allows to rigorously generalize
results obtained in $d=2$ 
(partly by heuristic arguments and direct 
calculations) and at the same time
it provides a novel insight  of the  algebraic-geometric structure
underlying the optimal quantum cloning problem.
\section{cloners}%%%%%%%%%%%%%%%%%%%%%%%%
In this section  some mathematical  aspects 
of quantum states and quantum transformations of a  $d$-dimensional 
quantum system will be discussed.
In particular the optimization problem of imperfect cloning will be 
formulated in geometric fashion.
\subsection{ The GB Representation}
Let $\cal H$ be a $d$-dimensional Hilbert space.
The set $\mbox{End}({\cal H})$ of linear operators over $\cal H$
can be endowed with a metric structure in several ways. 
In view of its direct connection with the geometrical framework
of this paper, we shall consider $\mbox{End}({\cal H})$ as a metric space
with distance 
\begin{equation}
d(A,\,B)=2^{-1/2} \sqrt{(A-B,\,A^\dagger-B^\dagger)}
\label{metric}
\end{equation}
induced by the Hilbert-Schmidt scalar product $(A,\,B)\equiv \mbox{tr} A\,B^\dagger$
(the normalization has been chosen for later convenience).   
The  Lie algebra of hermitian $d\times d$ traceless matrices, $su(d)$, 
is a $D$-dimensional {\it real} subspace of $\mbox{End}({\cal H})$, where $D=d^2-1.$
One can choose a basis  $\{ {{\tau}}_i\}_{i=1}^D$ 
of $su(d)$  satisfying the relations
$( {{\tau}}_i,\, {{\tau}}_j) = 2\,\delta_{ij}.
$
The set ${\cal B}_1$ of the unit-trace Hermitean operators is a $D$-dimensional
hyperplane of $\mbox{End}({\cal H})$.
Any element  of $ {\cal B}_1$ can be written as 
\begin{equation}
\rho ({{\lambda}})= \frac{1}{d}\, {\bf{I}} + \frac{1}{2}
\,\sum_{i=1}^D \lambda_i {{\tau}}_i
\label{Bloch}
\end{equation}
The vector ${{\lambda}}\equiv(\lambda_1,\ldots,\lambda_D)\in{\bf{R}}^D$
will be referred to as the {\it Generalized Bloch Representation} [GBR] of $\rho.$
Equation (\ref{Bloch}) defines a mapping $m\colon {\cal B}_1\rightarrow {\bf{R}}^D$
that associates to any $\rho\in{\cal B}_1$ its GBR vector,  
such that $\rho=\rho(m(\rho)).$
Let ${\cal P}\subset {\cal B}_1$
the set of pure states on ${\cal H},$ and ${\cal S}=hull({\cal P})$ its convex hull 
(the {\it state space}).The corresponding objects over ${\cal H}^{\otimes\,N}$
will  be denoted by same notation with an extra index $N.$\\
In the following ${\bf{R}}^D$ will be considered endowed with the  geometrical
structure associated with  the {\it euclidean} scalar product 
$\langle x,\,y\rangle\equiv \sum_{i=1}^D x_i\,y_i$, and norm $\|x\|\equiv 
\sqrt{\langle x,\,x\rangle}.$ \\
Let $S_{D-1}\subset {\bf{R}}^D$ be the $(D-1)$-dimensional hypersphere
with radius $ R_d \equiv\sqrt{ 2\,(1-1/d)},$ and
$B_{D}$  the
ball $\partial B_{D}=S_{D-1}.$ 
For $d=2$ (the {\it qubit} case) one finds $R_2=1$; $B_2$ is the Bloch sphere.
The basic properties of the GBR mapping $m$ are collected in the following\\
{\bf{Proposition 1}}
i) $m$ is an affine bijection; 
ii) $(\sigma,\,\rho) =d^{-1}+2^{-1} \langle m(\rho),\,m(\sigma)\rangle,$
and $d(\rho,\,\sigma)=2^{-1}\,\|m(\rho)-m(\sigma)\|$; 
iii) $m({\cal P}) \subset S_{D-1}$, and
$m({\cal S}) \subset B_D$.\\
{\it Proof}\\
i) In order to prove affinity one has to check 
that $m(\mu\rho_1+(1-\mu)\,\rho_2)= \mu\, m(\rho_1)+ (1-\mu)\,\rho_2,
\forall \rho_1,\,\rho_2\in{\cal S},\, 0\leq\mu\leq 1.$
Since the components of $m$ are given by $m_i(\rho) =(\tau_i,\,\rho),\,(i=1,\ldots,D)$
this is immediate. Bijectivity follows from the next point.
ii) Derives by straightforward calculation using orthogonality
of the $\tau_i$'s.. 
iii) If $\rho\in{\cal P}$ one has $\rho^2=\rho$, 
then (by previous point) $1=\mbox{tr} \rho^2=1/d+ 1/2\,\|\lambda\|^2$, 
whence $m(\rho)\in S_{D-1}.$
For general states of ${\cal S} $ one has  -- due to affinity of $m$ --
$m({\cal S})= m( hull({\cal P}))= hull (m({\cal P}))\subset hull (S_{D-1})= B_D$. 
$\hfill\Box$\\
{\it It  is important to notice  that, for $d>2,$ $m({\cal S})$ is a  proper subset 
 of $B_D.$}\\
Indeed: if $\rho\in{\cal S}\Rightarrow \mbox{tr} (\sigma\,\rho)\ge0,\,
\forall \sigma\in{\cal P},$
but $\mbox{tr} (\sigma\,\rho)=1/d+ 1/2\,\langle m(\sigma),\,m(\rho)\rangle,$
then $\|m(\sigma)\|\,\|m(\rho)\| \cos \theta\equiv
\langle m(\sigma),\,m(\rho)\rangle \ge -2\,d^{-1},$
and since $\|m(\rho)\|\le \|m(\sigma)\|= 2\,(1-1/d)$ one has
$ \cos\theta\ge (1-d)^{-1}.$
This constraint is automatically  fulfilled for all the 
elements of $B_D$
just for $d=2.$
In the general case one has a maximum allowed 'angle' $\theta_M(d)=\cos^{-1}[1/(1-d)]$ 
 Notice that  $\theta_M(\infty)=\pi/2.$                 
For example suppose $\rho(\lambda)\in{\cal P}$: then 
$(\rho(\lambda),\,\rho(-\lambda))=d^{-1}-2^{-1}\,R_d^2,$
a quantity which is non-negative just for $d\le 2.$
\\
We recall that: i) A  mapping $T\colon {\cal B}_1\mapsto{\cal B}_1$
is referred to as {\em positive} if $T({\cal S})\subset {\cal S}$
i.e., it preserves positivity ;
ii) An {\em affine}  mapping $T\colon {\cal B}_1\mapsto{\cal B}_1$
is referred to as {\em completely positive}  if $\forall n$
the (trivially extended) maps 
\begin{equation}
T_n\colon {\cal B}_1\otimes \mbox{End}({\bf {C}}^n)
\mapsto {\cal B}_1\otimes \mbox{End}({\bf {C}}^n)
 \end{equation} 
given by $T_n=T\otimes{\bf{Id}}$ are positive \cite{KRA}.
The set of positive [completely positive]
maps of a subset $X\subset {\cal B}_1$ into itself,
will be denoted by ${\cal M}(X)$ 
[$CP(X)$].

The GBR  can be naturally lifted to the space  ${\cal M}({\cal B}_1)$  of 
(not necessarily affine) {\it positive} mappings of ${\cal B}_1$ into itself
by the formula $ T\rightarrow {\cal T}= m\circ T \circ m^{-1},$ 
or, equivalently, by  the following commutative diagram
\\
\def\normalbaselines{\baselineskip20pt\lineskip3pt\lineskiplimit3pt}
\def\mapright#1{\smash{\mathop{\longrightarrow}\limits^{#1}}}
\def\mapdown#1{\Big\downarrow\rlap{$\vcenter{\hbox{$\scriptstyle#1$}}$}}
\begin{eqnarray}
\matrix{{\cal B}_1 & \mapright{T} & {\cal B}_1 \cr
        \mapdown{m} && \mapdown{m} \cr
        {\bf{R}}^D & \mapright{ {\cal T} } & {\bf{R}}^D &}
        \nonumber
\end{eqnarray}
The next proposition shows that
 $CP$-maps will be associated with {\it affine transformations} in ${\bf{R}}^D.$\\
{\bf{Proposition 2}}
Let $T\in{\cal M}( {\cal B}_1)$ be a trace-preserving CP-map. Then: 
i) $T({\bf{I}}/d)={\bf{I}}/d+
\sum_j c_j {{\tau_j}}$; 
ii) $T({{\tau_i}}) = \sum_{j=1}^D
 M_{ji} {{\tau_j}}.$\\
{\it{Proof}}\\
i) $T({\bf{I}}/d)$ must be a trace one hermitian operator by definition of CP-map.
ii) $T({{\tau_i}})$ must be traceless and hermitian; the statement
 follows from the fact that $\{ \tau_i\}$ are a $su(d)$ basis.\\
Therefore, if $\rho$ has the form (\ref{Bloch}), 
\begin{eqnarray}
T(\rho) &=& \frac{1}{d}\, {\bf{I}} +\frac{1}{2}
\,\sum_{i, j=1}^D \lambda_i M_{ji} \tau_j \nonumber\\
&=& \frac{1}{d}\,{\bf{I}}+ \frac{1}{2}\,\sum_{j=1}^D \lambda^\prime_j \,\tau_j, 
\label {Tbloch}
\end{eqnarray}
where $ \lambda^\prime = {\bf{M}}(\lambda)+ {\bf{c}},
\, {\bf{M}}= ( M_{ij} )\in\mbox{End}({\bf{R}}^D),\, {\bf{c}}\in{\bf{R}}^D.$ $\hfill\Box$\\
{\it This realizes an (affine) mapping $M$ 
between the trace-preserving CP-maps
on ${\cal B}_1$ and the affine transformations of  ${\bf{R}}^D$ in itself.
$M\colon CP({\cal B}_1)\rightarrow {\mbox{Aff}}({\bf{R}}^D)
\colon T\rightarrow {M}(T)= m\circ T \circ m^{-1}.$ }\\
A particularly relevant class of $CP$-maps is given by the unitary transformations. 
Any $X\in SU(d)$ defines, via the adjoint action, a CP-map on 
${\cal B}_1,\;\rho\rightarrow \mbox{Ad} X(\rho)\equiv X\,\rho\,X^\dagger.$
The following proposition shows that unitary transformations correspond, 
in the GBR, to rotations.
\\
{\bf{Proposition 3}}
{\it $\varphi\equiv M
\circ {Ad}$ is a homomorhism of $SU(d)$ in $SO(D).$ 
}
\\
{\it Proof}\\
First observe that from $\mbox{Ad}X({\bf{I}})={\bf{I}},\,(\forall X\in SU(d))$
there follows that ${\bf{c}}=0:$ {\it $M(T)$ is linear.}  
Since obviously $M(T_1\,T_2)= {M}(T_1)\,{M}(T_2)$, one has just
to check that, 
for any $X\in SU(d),$ the mapping
$\lambda\rightarrow m( \mbox{Ad} X(m^{-1}(\lambda)))$
preserves scalar product (and then the norm)  on ${\bf{R}}^D.$
Indeed $ \langle \lambda,\,\mu\rangle =2\,(\mbox{tr} [\rho(\lambda)\,\rho(\mu)]
- 1/d),$
and the trace is  Ad-invariant.
Since $X\,\tau_i\, X^\dagger= \sum_j X_{ji} \tau_j$, one has
that the induced ${\bf{R}}^D$ mapping has the form $\lambda\rightarrow {\bf{X}}(
\lambda)$
where the matrices ${\bf{X}} =(X_{ji} )$ are the adjoint representatives of $SU(
d).$ $\hfill\Box$
\\
Since $SU(d)$ acts (via Ad) {\it transitively} \cite{notTran}
 on $\cal P$, it follows immediately
that the subgroup $\varphi(SU(d))$ acts transitively over $m({\cal P}).$\\
{\it Once again it is worth emphasizing that, for $d> 2,$ 
$\varphi(SU(d))$ is a { proper} subset of $SO(D)$.}
This can be easily understood observing that any pair $\lambda,\,\mu$ of points of
$S_{D-1}$ are connected by an orthogonal transformation $R_{\lambda, \mu}$;  
in particular one can have $\lambda\in m({\cal P})$ and $\mu\not\in m({\cal P}).$   
Since $m({\cal P})$ is $SU(d)$-invariant, $R_{\lambda, \mu}\not\in \varphi(SU(d)).$
\subsection{Optimality}
The metric structure over $\cal S$ allows us to introduce several
natural 'figures of merit' for cloning.
For example let us consider, for given $T\in{\cal M}({\cal B}_1),$
the functional 
$F_1\colon {\cal S}\rightarrow [0,\,1]$ given by
\begin{eqnarray}
F_1(T,\,\rho) &=&  1-\left [ d(\rho,\,T(\rho))\right ]^2 \nonumber \\
&=&  2^{-1} \,\delta(T(\rho)) + F(T,\,\rho),
\label{fidelity}
\end{eqnarray}
where
$\delta(T(\rho))\equiv 1- \mbox{tr}\, T^2(\rho)$ is the {\it idempotency deficit} (or linear entropy)
of $T(\rho)$ and  
$F(T,\,\rho) \equiv \left (T(\rho),\,\rho\right ),$
is the {\it pure state fidelity} \cite{FID}.
\\
The naturality of $F_1$ as (state dependent) measure of cloning goodness should be clear:
it is maximum (equal to $1$) when $\rho=T(\rho)$ and minimum ($0$) 
when $\rho$ and $T(\rho)$
have disjoint supports. Moreover both contributions $\delta$ and $F$
to the 'merit' function $F_1$ have a clear geometrical meaning in ${\bf{R}}^D$. 
Indeed,
by using the GBR one finds (from Proposition 1)
\begin{eqnarray}
\delta(T(\rho)) &=& \frac{1}{2}\,(R_d^2-\|{\cal T}(\lambda)\|^2), \nonumber\\
F(T,\,\rho) &=& \frac{1}{d} + \frac{1}{2}\,\langle{\cal T}(\lambda),\,\lambda\rangle.
\end{eqnarray}
It is interesting to consider 
a special class of transformations for which the quality
of the cloning process is independent on the (pure) input state \cite{BRDI}.
This motivates the following\\
{\bf{Definition 1}}
A map $T\in{\cal M}({\cal S})$ is {\it universal} if $F_1(T,\rho)$ is independent
on $\rho\in{\cal P}$.\\
For general maps (i.e. non universal)  one can be
 interested in optimizing the worst case, with {\it pure} initial input.
Therefore it is natural to introduce the quantity
\begin{equation}
\tilde F_1(T) =\min{\rho\in{\cal P}}\,F_1(T,\,\rho). 
\end{equation}
The following proposition will turn to be useful:\\
{\bf Proposition 4} i) {\it $\tilde F_1$ is a concave functional over ${\cal M}({\cal S}).$}
ii) If $U\in SU(d)$ and $T_U\in {\cal M}({\cal S})$ is defined by
$T_U(\rho)=U^\dagger\,T(U\,\rho\,U^\dagger)\,U$, one has $\tilde F_1(T_U)= \tilde F_1(T).$
\\
{\it Proof}\\
i) 
Let $T_1,\, T_2\in{\cal M}({\cal S}), \mu\in{\bf{R}}^+_0.$ In view of the concavity of $\delta$
one has 
$
F_1(\mu\,T_1+(1-\mu)\,T_2,\,\rho)\ge 2^{-1}\,\mu\,\delta(T_1(\rho))+2^{-1}\,(1-\mu)\,
\delta (T_2(\rho))
+\mu\,F_1(T_1,\,\rho)+(1-\mu)\,F_1(T_2,\,\rho)$. Then, 
by  the superadditivity of the infimum
one gets
$$
 \tilde F_1(  \mu\,T_1+(1-\mu)\,T_2)\ge
\mu\,\tilde F_1(T_1) + (1-\mu)\,\tilde F_1(T_2).
$$
ii) Explicitly using $SU(d)$-invariance of the metric, and transitivity of $SU(d)$-action 
over $\cal P$, 
\begin{eqnarray} 
\tilde F_1(T_U)&=&\inf{\rho\in{\cal S}} F(T_U,\,\rho)= \nonumber \\
& &\inf{\rho\in{\cal S}}
(1-d^2(\rho,\, U^\dagger\,T(U\,\rho\,U^\dagger)\,U)) \nonumber \\ 
&=&\inf{\sigma\in{\cal S}}
(1-d^2(U^\dagger\,\sigma\,U,\, U^\dagger\,T(\sigma)\,U))\nonumber \\ 
&=&\inf{\sigma\in{\cal S}}
(1-d^2(\sigma,\,T(\sigma))=\tilde F_1(T). \nonumber 
\end{eqnarray} 
$\hfill\Box$\\
The mapping $T\rightarrow T_U$ defines a $SU(d)$-action $\Phi$ such that
$T_U\equiv \Phi(U,\,T)$
 over ${\cal M}({\cal S})$. 
Point ii) of the previous proposition simply states that the 
quality of cloning is constant along the 
orbits of $\Phi$.
The fixed points of $\Phi$ therefore play a special role.  
\\
{\bf{Definition 2}}
A map $T\in{\cal M}({\cal S})$ is
{\it covariant} iff $T_X=T,\;
\forall  X\in SU(d).$\\
Next proposition shows that covariance implies universality
and imposes strong geometrical constraints to the GBR.\\
{\bf{Proposition 5}} {\it Suppose $T\in{\cal M}({\cal S})$
covariant. Then: i) $T$ is universal, 
\,ii) ${\bf{U}}\,{\cal T}_k(\lambda)= {\cal T}_k({\bf{U}}\,\lambda),$
$\forall \lambda\in {\bf{R}}^D,\; {\bf{U}}\in\varphi (SU(d))$;  
iii) $\|{\cal T}(\lambda)\|$ and 
$\langle {\cal T}(\lambda),\,\lambda\rangle$
are  constant 
over $m({\cal P})$
}
\\{\it Proof}\\ 
i)  Since $\mbox{Ad} SU(d)$ is transitive over $\cal P$, it suffices to show that
$F_1(T,\rho)= F_1(T,\mbox{Ad} X(\rho)),\,(\forall X\in SU(d),\,\rho\in{\cal P}).$
Indeed, 
$\delta (T(\mbox{Ad} X \rho))= \delta( \mbox{Ad} X\,T(\rho))=\delta(T(\rho)),
$
and
\begin{eqnarray}
 & &F(\rho,T)=\mbox{tr}\left ( \rho\,T(\rho)\right )=
\mbox{tr}\left ( X^\dagger\,\rho\,T(\rho)\, X\right ) =
\nonumber \\
& &\mbox{tr}\left ( X^\dagger\,\rho\,X\,T(X^\dagger\,\rho\, X) \right )
=\mbox{tr}\left ( \mbox{Ad}X(\rho)\,T(\mbox{Ad}X(\rho))\right ). 
\end{eqnarray}
ii) This point requires just an explicit check.
iii) 
If $\lambda,\,\lambda^\prime\in m({\cal P})
\Rightarrow \exists {\bf{U}}\in \varphi(SU(d)),\;{\it s.t.}\; 
{\bf{U}}\lambda=\lambda^\prime.$ Then $ \|{\cal T}(\lambda^\prime)\|=
\|{\cal T}({\bf{U}}\lambda)\|=\|{\bf{U}}{\cal T}(\lambda)\|=
\|{\cal T}(\lambda)\|.$
Moreover, $\forall \lambda,\lambda^\prime\in m({\cal P})$, 
\begin{eqnarray}
& &\langle {\cal T}(\lambda),\,\lambda\rangle = \langle {\bf{U}}{\cal T}(\lambda),\,{\bf{U}}
\lambda\rangle = \nonumber \\
 & & \langle {\cal T}({\bf{U}}\lambda),\,{\bf{U}}
\lambda\rangle
 = \langle {\cal T}(\lambda^\prime),\,
\lambda^\prime\rangle.
\end{eqnarray}
$\hfill\Box$\\
Mappings satisfying relation ii), for
${\bf{U}}$ belonging to some group $\cal G$,  are known as {\it $\cal G$-automorphic 
functions}.  Therefore point ii) of the previous proposition can be rephrased 
saying that {\it GBR of covariant maps of ${\cal M}({\cal S})$ are $\cal G$-automorphic 
functions of ${\bf{R}}^D$ in itself, where ${\cal G}\equiv\varphi(SU(d)).$}\\
Of course any linear mapping $M\in\mbox{End}({\bf{R}}^D)$ is $\cal G$-automorphic for {\it any} subgroup 
${\cal G}\subset GL(D,\,{\bf{R}})$ such that $[M,\,{\cal G}]=0$
($M$ belongs to the {\it centralizer} of $\cal G$).
An example of $SO(D)$-automorphic functions is given by 
${\cal T}(\lambda) =f(\|\lambda\|)\,\lambda$, 
with $f\colon {\bf{R}}\rightarrow (0,\,1)$ .
Notice that, for these mappings,  the functions  (\ref{fidelity})
are  constants over $D-1$-dimensional spheres.\\

Let ${\cal M}^\prime$ a convex $\Phi$-invariantt subset of ${\cal M}({\cal S})$
The notion of optimality  used in this paper  is given by\\
{\bf{Definition 3}}
 Let ${\cal M}^\prime$ a convex $\Phi$-invariantt subset of ${\cal M}({\cal S}).$
A map $T^*\in{\cal M}^\prime$ is {\it optimal} ( in ${\cal M}^\prime$) if 
\begin{equation}
\tilde F_1(T^*)=\sup{T\in {\cal M}^\prime} \tilde F_1(T).
\end{equation}
Now we show that, {\it as far as optimality is concerned, one can restrict oneself
to covariant transformations without loss of generality.}
The basic idea  is very simple:
since our 'merit' functional $F_1$ is concave and $SU(d)$-invariant
one can be build, for any given $SU(d)$-orbit,  a convex average transformation $T^*$
non-decreasing   the cloning quality (i.e. $\tilde F_1(T^*)\ge \tilde F_1(T)$).
$T^*$ will be, by construction, covariant
and it is clear that 
the one associated with an optimal cloner will turn out to be optimal as well.
\\ 
{\bf{Proposition 6}}
{\it The optimal map can be chosen to be covariant}.\\ 
{\it Proof}\\
Given the  $SU(d)$ action $\Phi$ over ${\cal M}^\prime,$ the element 
of ${\cal M}^\prime$,
 $T^*=\int_{SU(d)}  d\mu(X) T_X$ (where $T_X = \Phi (X,T)$, for a 
given $T\in {\cal M}^\prime$) is a covariant  map. 
Indeed, for any $Y\in SU(d)$, $ T^*_Y=\int_{SU(d)}  d\mu(X) (T_X)_Y=
\int_{SU(d)}  d\mu(X) T_{XY}=\int_{SU(d)}  d\mu(XY) T_{XY}=T^*,$
where the invariance of the Haar measure $d\mu$ was used \cite{CORN}.
If $T$ is optimal $\tilde F_1(T)\ge \tilde F_1(T^*)\ge \int_{SU(d)}  d\mu(X)
\tilde F_1(T_X)=\tilde F_1(T)$; thus $\tilde F_1(T^*)=\tilde F_1(T)$ 
(where we used the concavity of $\tilde F_1,$  Proposition 4, and the normalization
$\int_{SU(d)}  d\mu(X)=1$ of the measure $d\mu$). $\hfill\Box$\\
Notice that if $T$ is universal, then the mapping $T^*$ introduced in Proposition 6 
has the same value of the merit functional.
Indeed $\forall \rho\in{\cal P}$ one has  
\begin{eqnarray}
& &\tilde F_1(T) =  F_1(T,\,\rho)=
\int d\mu(X) F_1(T,\,X\,\rho\,X^\dagger )\nonumber \\
&=&\int d\mu(X) (1-d^2(X\,\rho\,X^\dagger,\,T(X\,\rho\,X^\dagger)))\nonumber \\
&=&
\int d\mu(X) (1-d^2(\rho,X^\dagger\,T(X\,\rho\,X^\dagger)\,X))\nonumber \\
&=& 1-d^2(\rho,\,T^*(\rho))= \tilde F_1(T^*), 
\end{eqnarray}
where we have used linearity in $T$ of $F_1,$ normalization of $d\mu$ 
and $SU(d)$-invariance of the metric.\\
This observation makes clear that for optimization purposes 
t one can identify the notion of covariance and universality:
{\it a covariant map is universal and for any universal map there exists
a covariant map with same cloning quality.}\\ 
The next theorem shows that the structure of {\it affine}
covariant maps is very simple.\\
{\bf{Proposition 7}}
{\it $T\in CP({\cal S})$ is covariant   iff  ${\cal T} =\xi\, {\bf{I}}$
with $\xi\in (0,\,1)$.}\\
{\it Proof}\\
a) If ${\cal T}=\xi\,{\bf{I}}$, it is trivial to check that
$T$ is covariant. 
b) The components of the GBR of $T$ are given by 
${\cal T}_i(m(\rho))= (\tau_i,\,T(\rho))=( F_i,\,\rho),$
where $F_i= T^t(\tau_i)$ are traceless operators 
($T^t$ is the transpose map of $T$ with respect to 
the Hilbert-Schmidt scalar product). 
On has to show that $F_i=\xi\,\tau_i.$
If $T$ is covariant, 
$T(X\,\rho\,X^\dagger)=X\,T(\rho)\,X^\dagger$; therefore 
\begin{eqnarray}
(\tau_i,\, T(X\,\rho\,X^\dagger)) &=&
(F_i,\,X\,\rho\,X^\dagger)= (X^\dagger\,F_i\,X,\,\rho)=\nonumber \\
(\tau_i,\,X\,T(\rho)\,X^\dagger) &=& (X^\dagger\,\tau_i\,X,\, T(\rho)) =\nonumber \\
\sum_j X_{ji}\,( \tau_j,\,T(\rho)) &=& \sum_j X_{ji}\,( F_j,\,\rho). 
\end{eqnarray}
Since this equality holds for any $\rho\in{\cal S}$, one gets
\begin{equation}
X^\dagger\,F_i\,X=\sum_j X_{ji}\, F_j,\label{transf}
\label{adj}
\end{equation}
namely the $F_i$'s transform under the adjoint action of $SU(d)$ as the $\tau_i$'s. 
As such action is {\it irreducible}, this implies that $F_i=\xi\,\tau_i,\,(i=1,\ldots,D).$
Indeed let $ F_i= \sum_j M_{ij}\,\tau_j,$: from equation (\ref{adj}) one finds
$[{\bf{X}},\,{\bf{M}}]=0,\,\forall X\in SU(d),$
then -- by Schur's lemma -- ${\bf{M}}=\xi\,{\bf{I}}.$
Moreover $\xi\in[0,\,1]$ due to positivity requirement.
$\hfill\Box$ \\
 A covariant map $T\in CP({\cal S})$ has the form \cite{BREK}
given by 
\begin{equation}
T(\rho) =(1-\xi)\,d^{-1}\,{\bf{I}}+\xi\,\rho.\label{BCE}
\end{equation}  
The following  example shows that one can  
have a whole family
of covariant, positive, trace-preserving, {\it non-linear} maps of ${\cal S}$
in  itself
$
T_\Gamma(\rho) =(1-\Gamma[\rho])\,d^{-1}\,{\bf{I}} +
\Gamma[\rho]\,\rho,
$
where $\Gamma\colon {\cal S}\rightarrow (0,\,1)$
is a $SU(d)$-invariant non-linear functional.
Such functionals can be built, for example,
given any non-linear map $\gamma\colon (0,\,1)\rightarrow (0,\,1)$
by any convex superposition of the
maps $\Gamma_n(\rho)=\gamma(\mbox{tr} \,\rho^n).$
\\
{\it
These maps, restricted to $\cal P$, amount to a simple (state independent)
shrinking of the generalized Bloch vector $m(\rho).$
Nevertheless, since they are not affine, the property 
cannot be extended to the whole $\cal S.$
}
\section{Cloners $N\rightarrow M$}%%%%%%%%%%%%%%%%%%%%%%%%%%%%%%%%%%%%%%%%%
Now we turn the  $N\mapsto M$ cloning. 
In this section we shall set $\tau_0\equiv{\bf{I}},\,\lambda_0\equiv d^{-1}$, 
and $\lambda_i\mapsto\lambda_i/2.$ 
Let us consider the $N$-system state 
$\rho_{N}\equiv\rho(\lambda)^{\otimes\,N}$, 
\begin{equation}
\rho_{N}=\sum_{i_1,\ldots,i_N=0}^D \lambda_{i_1}\cdots
\lambda_{i_N} \,\otimes_{k=1}^N\tau_{i_k}
=\sum_{ {\bf{i}}\in{\cal F}_{N,D} }\lambda_{\bf{i}}\,\tau_{\bf{i}}, 
\label{input}
\end{equation}
where ${\cal F}_{N,D}$ is the set of the maps from $\{0,\ldots,N\}$
to $\{0,\ldots,D\},$ and $\lambda_{\bf{i}}\equiv\prod_{k=0}^N \lambda_{i_k},\,
\tau_{\bf{i}}\equiv \otimes_{k=0}^N\tau_{i_k}.$
Notice that in equation (\ref{input}) the only non trace-less
term is $\lambda_{\bf{0}}\,\tau_{\bf{0}}\equiv d^{-N}\,{\bf{I}}^{\otimes\,N}.$\\
The set of trace-preserving $CP$-maps  from  ${\cal S}^N$ to ${\cal S}^M$
  will be denoted as $CP_{M,N}.$\\
{\it The problem is now to find the optimal (with respect to some 
some criterion) transformations 
of $CP_{M,N}.$}\\
Since $X\in SU(d)$ acts naturally over  $CP_{M,N}$ by 
$\Phi^N\colon (X,\,T) \mapsto T_X$
in which
\begin{equation}
T_X(\rho)= 
\mbox{Ad}^{\otimes\,M} X^\dagger (\,T(\mbox{Ad}^{\otimes\,N} X (\rho))), 
\end{equation} 
the   notion of covariance is immediately extended to $CP_{M,N}.$
It means that $T$ 'intertwines' between the $N$ and $M$-fold
tensor representations of $SU(d)$.  This can be pictorially
described by the following commutative diagram
\def\normalbaselines{\baselineskip20pt\lineskip3pt\lineskiplimit3pt}
\def\mapright#1{\smash{\mathop{\longrightarrow}\limits^{#1}}}
\def\mapdown#1{\Big\downarrow\rlap{$\vcenter{\hbox{$\scriptstyle#1$}}$}}
\begin{eqnarray}
\matrix{{\cal S}^N & \mapright{\mbox{Ad}^{\otimes\,N} \, X } & {\cal S}^N \cr
        \mapdown{T} && \mapdown{T} \cr
        {\cal S}^M & \mapright{\mbox{Ad}^{\otimes\,M}\, X^\dagger } & {\cal S}^M }
       \nonumber
\end{eqnarray}
To grasp what covariance means consider
a set of operators $\{\phi_i\}_i$ in the domain of 
$T\in CP_{M,N},$ 
that under the $\mbox{Ad}^{\otimes\,N}$-action of $SU(d)$
transform according to an {irreducible} representations ${\bf{R}}$
(i.e. $ \mbox{Ad}^{\otimes\,N} X (\phi_i) =\sum_j {\bf{R}}_{ji}(X) \phi_j$).
If $T$ is covariant then 
$\tilde \phi_i\equiv T(\phi_i)$ transform
under $\mbox{Ad}^{\otimes\,M}$ according the {\it same} irrep.
{\it 
In other words a covariant mapping conserves the $SU(d)$ symmetry content of the states.}
For example $ [X^{\otimes\,N},\,\rho]=0
\Rightarrow  [X^{\otimes\,M},\,T(\rho)]=0,$ in particular
if $\rho=d^{-1}\,{\bf{I}}$ one has that $T(\rho)$
belongs to the centralizer ${\cal A}_{d,M}$ of the $n$-fold tensor
representation of $SU(d).$
${\cal A}_{d,M}$ is an algebra generated by the representatives
of the  symmetric group $S_N$ acting in the natural way.
Of course for ${\cal A}_{d,1}\propto {\bf{I}}.$\\
In the multi-system case now under consideration, one
has also the natural action of the symmetric group $S_M$ over ${\cal S}^M$
[if $\sigma\in S_M$ and $\rho =|\Psi\rangle\langle\Psi|$, $\sigma\cdot\rho\equiv U_\sigma|\Psi\rangle\langle\Psi|
U_\sigma^\dagger,$ where $U_\sigma\otimes_{j=1}^M |\psi_j\rangle=\otimes_{j=1}^M
|\psi_{\sigma(j)}\rangle.$]
therefore one can consider the maps
$T_\sigma(\rho) =\sigma\cdot T(\rho)\, (\sigma\in S_M).$\\
{\bf Definition 4}
A map $T\in CP_{M,N}$ such that $T\equiv T_\sigma,\, \forall \sigma\in S_M$
will be referred to as {\it symmetric}.\\
{\it Remark} For symmetric maps $T(\rho)$ is totally symmetric operator.
Let us denote with $ \mbox{tr}_{\bar k}$ the trace over all but the $k$-th factor 
of the tensor product ${\cal H}^{\otimes\,M}.$
One can associate, to any element  $T\in CP_{M,N}$, $M$ {\it reduced maps}
of ${\cal M}({\cal S})$ defined by the rule
$T_k\colon \rho\rightarrow \mbox{tr}_{\bar k}\, T(\rho^{\otimes\,N})$. \\
{\bf Proposition 8}
{\it   The maps
$\{ T_k\}_{k=1}^M$
fullfill the  following
i)  $T_k\in {\cal M}({\cal S})$.
ii) The GBR of the $T_k$'s have components that are polymomials of order $N.$
iii) If $T$ is symmetric the $T_k$'s are identical.  
iv) If $T$ is covariant so are the $T_k$'s.}\\
{\it  Proof}\\
i) The $T_k$'s are positive in that they are compositions of the positive maps.
ii)
One has
$T(\tau_{\bf{i}})=\sum_{{\bf{j}}\in{\cal F}_{M,D}}
M_{{\bf{j}},{\bf{i}}}\,\tau_{\bf{j}},$
therefore
$
T(\rho_{N})=\sum_{{\bf{j}}\in{\cal F}_{M,D}}
\lambda_{\bf{j}}^\prime \tau_{\bf{j}},$
where $\lambda_{\bf{j}}^\prime= \sum_{{\bf{i}}\in{\cal F}_{N,D}}
M_{{\bf{j}},{\bf{i}}}\,\lambda_{\bf{i}}$.
In particular
$ T(\tau_{\bf{0}})=d^{-M}\,{\bf{I}}^{\otimes\,M}+
\sum_{{\bf{j}}\neq {\bf{0}} } c_{ {\bf{j}}} \tau_{\bf{j}},
$ and
${\bf{i}}\neq {\bf{0}}\Rightarrow \mbox{tr} T(\tau_{\bf{i}})=0
\Rightarrow M_{ {\bf{0}},{\bf{i}} }=0.$ 
Moreover $\mbox{tr}_{\bar k} \,\tau_{\bf{j}} =\tau_{j_k}\,d^{M-1}
\prod_{l\neq k}\delta_{j_l, 0}.$
Therefore
$
T_k(\rho_{N})=d^{-1}\,{\bf{I}}+\sum_{j=1}^D{\cal T}^j_k(\lambda)\, \tau_j,
$ where
\begin{equation}
{\cal T}_k^j(\lambda)
 =d^{M-1}\,\sum_{{\bf{i}}\neq {\bf{0}}} (M_{ {\bf{j}}_k,{\bf{i}} }
\lambda_{\bf{i}} +c_{{\bf{j}}_k}). \label{nl}
\end{equation}
Here ${\bf{j}}_k$ is a $M$-component vector with $j$ in the $k$-th entry
and zero elsewhere.
iii) If $T$ is symmetric, it is simple to check that 
$M_{{\bf{j}}\circ\sigma,{\bf{i}}}=M_{{\bf{j}},{\bf{i}}}$, and 
$c_{{\bf{j}}\circ\sigma}=c_{{\bf{j}}},\;
\forall \sigma \in S_M,\,
{\bf{i}}\in{\cal F}_{N,D},\, {\bf{j}}\in{\cal F}_{M,D}$.  
In particular, if $l,\,k\in\{1,\ldots,M\}$, by applying the transposition
$\sigma_{kl}=(k, l)$ one finds ${\cal T}(\lambda)_k^j={\cal T}(\lambda)_l^j.$
iv) One proves, by direct calculation, that 
\begin{eqnarray}
& &T_k( X\,\rho\, X^{\dagger})=
 \mbox{tr}_{\bar k} T( (X\,\rho\, X^{\dagger})^{\otimes\,N}) \nonumber \\
&=& \mbox{tr}_{\bar k} T( X^{\otimes\,M} \rho^{\otimes\,N}\, X^{\dagger\otimes\,M})=
\mbox{tr}_{\bar k} X^{\otimes\,M} T(\rho^{\otimes\,N})  X^{\dagger\otimes\,M} 
\nonumber \\
& &X\,\mbox{tr}_{\bar k} T(\rho^{\otimes\,N}) X^{\dagger}= X\,T_k(\rho)\,X^\dagger.
\end{eqnarray}
$\hfill\Box$\\
{\bf{Definition 5}} We introduce, for the elements of $CP_{M,N}$, the (global) figures of merit
based on the quality of the reduced clones
\begin{eqnarray}
F^{MN}_1 (T,\,\rho) &=& \min{k} F_1(T_k,\,\rho)
\;(\rho\in{\cal P}), \nonumber \\
\tilde F^{MN}_1(T)&\equiv& \min{\rho\in{\cal P}} F_1(T,\,\rho), 
\end{eqnarray}
the notion of optimality being given as for the reduced maps
for a convex, $\Phi^N$-invariant ${\cal M}^\prime_{M,N}
\subset CP_{M,N}$ .\\
The next proposition is an extension of proposition 6  to the $N\mapsto M$ case.\\ 
{\bf Proposition  9}
{\it An optimal  $T\in {\cal M}^\prime_{M,N}$ can be chosen covariant
and symmetric}.\\
{\it Proof}\\
Let us first observe that the functional $\tilde F_1^{MN}$ is constant over the orbits
of both the $SU(d)$ and $S_M$ actions. Indeed for 
$k=1,\ldots,M,\, U\in SU(d),\,\sigma\in S_M,$ one has
: i) $(T_k)_U=(T_U)_k$ ,from
 which $\tilde F_1^{MN}(T_U)=\tilde F_1^{MN}(T)$ and ii) $ (T_\sigma)_k= 
T_{\sigma^{-1}(k)}$,   
from which $\tilde F_1^{MN}(T_\sigma)=\tilde F_1^{MN}(T).$
Furthermore, it follows from linearity of the mapping $T\mapsto T_k,$ 
the properties of $\tilde F_1$,  
and $\min{k}$  that $\tilde F^{MN}_1$
is a concave functional over $CP_{M,N}.$
Now one can proceed as in Proposition 1, by  introducing the 
'covariantized' maps 
$ T^*_{\cal G}\equiv \int_{\cal G} d\mu(g)\, T_g \,
({\cal G}= SU(d),\, S_M$).
[For the symmetric group the covariant map associated
to $T$ is $T^*=(M!)^{-1}\sum_{\sigma\in S_M} T_\sigma$
]. $\hfill\Box$
\subsection{Universal Cloners}
Let us suppose that the map $T^{MN}\in CP_{MN}$ is  defined over the input set
\begin{equation}
{\cal S}_{in}\equiv \{\rho^{\otimes\,N},\,\rho\in{\cal P}\}. 
\end{equation}
According to Proposition 9, such a map can be assumed -- for optimality purposes --  
{\it covariant and symmetric.} 
The associated (reduced) pure-state fidelity, that has to be minimized over $m({\cal P}),$  is given
by equation (\ref{fidelity})  
[for a symmetric cloner $T^{MN}\in CP_{MN}$ we put $T^{MN}_k=T\, (k=1,\ldots,M)$, 
whereby ${\cal T}\colon  m({\cal P})\rightarrow m({\cal S})$]. \\
The next theorem shows how the deep geometrical meaning of covariance 
allows us to easily characterize 
the solutions  of the optimization problem.\\
{\bf Theorem 1}
{\it An optimal cloner $\rho\rightarrow  T^{MN}(\rho)
\,(\rho\in {\cal S}^N\bigcap\mbox{span}\,{\cal S}_{in}),$ 
 can be chosen in such a way that the associated reduced
map is given by a shrinking of the generalized Bloch vector.}\\
{\it Proof}\\
Due to the compacteness of $m({\cal P})$, there exists a $\lambda^*\in m({\cal P})$
such that $\tilde F(T)=1/4\,(R_d^2-\|{\cal T}(\lambda^*)\|^2)+
d^{-1}+1/2\,\langle {\cal T}(\lambda^*),\,\lambda^*\rangle$. 
Then 
\begin{eqnarray}
\tilde F(T) &\le& 1/4\,(R_d^2-\|{\cal T}(\lambda^*)\|^2)\nonumber \\  
 &+& d^{-1}+1/2\, \|\lambda^*\|\,\|{\cal T}(\lambda^*)\| . 
\end{eqnarray}
First notice that,  since $T$ can be chosen to be covariant, one has, 
from Proposition 5, that $\|{\cal T}(\lambda)\|$ is a constant 
over $m({\cal P})$.  Therefore: i) the first contribution to the fidelity does not depend on $\lambda,$
ii) the upper bound can be achieved if ${\cal T}(\lambda^*)=\xi \,\lambda^*.$
Now we observe that, as the  scalar product $\langle {\cal T}(\lambda),\,\lambda\rangle$
is constant over $m({\cal P})$ (Proposition 5), 
then 
${\cal T}(\lambda)=\xi(\lambda)\,\lambda.$
But  the automorphic constraint implies  
$\xi(\lambda)=\xi({\bf{U}}\lambda),\, \forall {\bf{U}}\in\varphi(SU(d))$
 whence -- by transitivity of the $SU(d)$-action over 
$m({\cal P})$ -- it must be $\xi\,|_{m({\cal P})}= \,const.$ 
The optimal (reduced) map has the  form
(\ref{BCE}). 
 Since this map is {\it affine}, 
 it can be extended to the whole
set of states belonging to the linear  span  of ${\cal S}_{in}.$ 
$\hfill\Box$\\
{\it {Remark}.} 
One must have 
$\mu_j^\prime=\xi \,\lambda_j\; (j=1,\ldots, D).$
Therefore $M_{j 0\ldots 0,{\bf{i}}}=0$ unless $\exists\, l\in\{0,\dots,N\}$
such that 
$i_m=0 \,(m\neq l)$ {\it and} $i_l=j.$
In this case one finds
\begin{equation}
\xi= \sum_{k=1}^N M(T)_{ {\bf{j}}_l,\,{\bf{i}}_k},\; (l=1,\ldots,M).
\label{factor} 
\end{equation}
Therefore
 \begin{equation}
T(\rho_{in})=d^{-M}\,{\bf{I}}^{\otimes\,M}+
N\,\xi\,\sum_{j=1}^D\lambda_j \Delta_M(\tau_i)+
 R(\lambda),
\end{equation}
where
$ \Delta_M(\tau_i)\equiv M^{-1}\sum_{l=1}^M\tau_i^{(l)}$
is the {\it coproduct} of $\tau_i$ 
[{\it i.e.} $\tau_j^{(l)}$ acts as $\tau_j$ in the $l$-th factor of the tensor product
and trivially in the others] and $ R(\lambda)$ contains all the tensor products
in which a factor $\tau_j\neq {\bf{I}}$ appears at least twice. 
$\hfill\Box$\\
\subsection{Algebraic approach}
In this section we shall show that the shrinking property (\ref{BCE}) 
follows from covariance alone.
To this aim it is convenient to turn to a more algebraic approach
in that the notion of covariance is naturally related to representation-theoretic concepts.
We consider now general $\rho\in{\cal S}.$\\
{\bf{Proposition 10}}
{\it The components of the map ${\cal T}$ are given by
${\cal T}_i(\lambda)=(F_i,\,\rho_\lambda^{\otimes\,N}),\,(i=1,\ldots,D)$
 where 
$F_i\in\mbox{End}({\cal H}^{\otimes\,N})$ are $S_N$-invariant, traceless hermitian operators.}\\
{\it Proof}\\
By using $S_M$-invariance of $T^{MN}$ one checks directly that the components of the 
map $\cal T$ are  
\begin{eqnarray}
 {\cal T}_i(\lambda) &=& (\tau_i,\,T(\rho))
=(\tau_i,\,\mbox{tr}_{\bar 1}
( T^{MN}(\rho_\lambda^{\otimes\,N}))) \nonumber \\
&= &\left ( \tau_i\otimes {\bf{I}}^{\otimes\,(M-1)},\, T^{MN}(\rho_\lambda^{\otimes\,N})\right )\nonumber \\
&=&\left ( \Delta_M(\tau_i),\, T^{MN}(\rho_\lambda^{\otimes\,N})\right )
= (F_i,\,\rho_\lambda^{\otimes\,N}),
\label{component}
\end{eqnarray}
where $ F_i\equiv 
{T^{MN}}^t(\Delta_M(\tau_i)).$
Now we observe that, since $\rho_\lambda^{\otimes\,N}$
is $S_N$-invariant, the $F_i$'s can be chosen to symmetric
\begin{eqnarray}
 (F_i,\,\rho_\lambda^{\otimes\,N})&=& \frac{1}{N!}\sum_{\sigma\in{S}_N}
(F_i,\, U_\sigma\,\rho_\lambda^{\otimes\,N} \,U_\sigma^\dagger)\nonumber \\
&=& \frac{1}{N!} \sum_{\sigma\in{S}_N}(U_\sigma^\dagger\,F_i\,U_\sigma,\,
\rho_\lambda^{\otimes\,N} )
= (\tilde F_i,\,\rho_\lambda^{\otimes\,N} ), 
\end{eqnarray}
where $\tilde F_i\equiv{1/M!}\sum_{\sigma\in{S}_N}U_\sigma^\dagger\,F_i\,U_\sigma$
is manifestly symmetric.
Tracelessness and hermiticity follow form the general properties of $CP$-maps.
$\hfill\Box$\\
From the covariance constraint it follows that 
\begin{equation}
( U^{\otimes\,N}\, F_i\, U^{\dagger\otimes\,N},\,\rho_\lambda^{\otimes\,N} )=
 \sum_{j=1}^D {\bf{X}}_{ji}(U)\,
( F_j,\,\rho_\lambda^{\otimes\,N} ). \label{27} 
\end{equation}
By introducing the functionals $\Lambda_\rho$ over $\mbox{End}({\cal H}^{\otimes\,N})$, 
$\Lambda_{\rho}\colon A\mapsto ( \rho,\,A)$,  equation (\ref{27}) can be rewritten as
$\Lambda_{\rho_N}(A^U_i)=0 \quad 
(\forall \rho\in {\cal S},\, U\in SU(d), \,i=1,\ldots,D),$ 
and
\begin{equation}
A_i^U\equiv U^{\otimes\,N}\, F_i\, U^{\dagger\otimes\,N}-\sum_{j=1}^D {\bf{X}}_{ji}(U) \,F_j.
\end{equation}  
Notice that, for $N=1,$ from (functional) equation $\Lambda_{\rho_N} (A^U_i)=0$   
follows the {\it operatorial}
equation (\ref{transf}). \\
Let us consider now the pure state case $\rho=|\psi\rangle\langle\psi|,\,
|\psi\rangle\in{\cal H}$.  
Let ${\cal H}^N_{sym}$the  totally symmetric subspace  of ${\cal H}^{\otimes\,N}.$
One has: i) ${\cal H}^N_{sym}$ is the space associated to the 
identity representation of $S_N$; 
ii) it is also the space of a totally symmetric 
(irreducible) representation $\phi_{s}$ of $SU(d)$;  iii)
${\cal H}^N_{sym}=\mbox{span}\{ |\psi\rangle^{\otimes\,N}\,\colon\,|\psi\rangle\in{\cal H}
\}.$ \\
{\bf{Theorem 2}}{\it
A covariant cloner over ${\cal H}^N_{sym}$ induces a mapping between reduced states given by
a simple shrinking of the generalized Bloch vector.}\\
{\it Proof}\\
It follows from i)--iii) that {\it the linear span
 of the operators $\rho^{\otimes\,N}$ ($\rho\in{\cal P}$)
is the  space  of states with support in
  ${\cal H}^N_{sym}$} \cite{WER}. 
In this case --  since  a symmetric operator leaves
${\cal H}^N_{sym}$ invariant --
 the $A_i^U$'s  can be considered as belonging to $\mbox{End}({\cal H}^N_{sym}).$
 The functional equations $\Lambda_{\rho}(A_i^U)=0$ 
then imply  the  operatorial equations $A_i^U=0$ over ${\cal H}^N_{sym}.$
Let  $\Phi$ be the representation over $\mbox{End}({\cal H}^N_{sym})$ associated with $\phi_s.$
Since $ \mbox{End}({\cal H}^N_{sym})\cong {\cal H}^N_{sym}\otimes
{\cal H}^{N*}_{sym}$ one has $\Phi\cong \phi_s\otimes\phi_s^*,$ the tensor product or two 
totally symmetric $SU(d)$-irreps, therefore
in the decomposition of $\Phi$ 
{\it each $SU(d)$-irrep appears once} \cite{CORN}.
As $A_i^U=0$ simply means that the $F_i$'s transform according 
to the adjoint representation, 
one must have $F_i\equiv
{T^{MN}}^t(\Delta_M(\tau_i))=\xi\, \Delta_N(\tau_i).$
Form this relation it follows 
(see equation (\ref{component})) that  
${\cal T}_i(\lambda)=\xi\,\lambda_i\,(i=1,\ldots,D).$
$\hfill\Box$\\
This proof helps to 
shed some light on the basic difference between the pure and the general (mixed)
state problem.
Let ${\cal H}^{\otimes\,N}=\oplus_{j\in{\cal J}} {\cal H}^{(j)}$ denote 
the decomposition of the input Hilbert space into $S_N$-isotopical components
(i.e. ${\cal H}^{(j)}$ is the subspace of vectors transforming according
a given $S_N$-irrep labelled by $j$).  
If $\Pi_j$ denotes the projector over ${\cal H}^{(j)}$
one has, for general $\rho$ that 
$\rho^{\otimes\,N} =\sum_{j\in{\cal J}}\lambda_j \rho^{(j)}_N,$
where $\rho^{(j)}_N\equiv \lambda_j^{-1}\, \Pi_j\rho^{\otimes\,N}\Pi_j,\,\lambda_j\equiv 
\mbox{tr}( \rho^{\otimes\,N} \,\Pi_j).$
In this case the relevant functional equations from covariance are
\begin{equation}
\sum_{j\in{\cal J}} \mbox{tr} ( \rho^{(j)}_N\,A_i^U  )=0,
\label{Func}
\end{equation}
$i=1,\ldots,D,\;U\in SU(d),$
where in each term $A_i^U$ can be considered as belonging to 
$\mbox{End}({\cal H}^{(j)}).$
When $\rho\in{\cal P}$ only the $j=0$ (${\cal H}^{(0)}\equiv{\cal H}^N_{sym}$)
term survives, and one succeeds in getting an operatorial equation.
In general one has to deal directly with equations (\ref{Func}), 
that represent a much weaker constraint on the cloner structure.\\
The next (almost obvious) corollary shows that concatenating optimal cloners
the shrinking factors multiply\\
{\bf Corollary 1}  %%%%%%%%%%%%
{\it Let $T_{1}\in CP_{M,N}$ and $T_{2}\in CP_{R,M}$ be 
symmetric and covariant maps.
Then: i)  $T_{2}\circ T_{1}$ is a covariant and symmetric map.
ii) Let $ r(T)$ denote the unique map of ${\cal M}({\cal S})$
associated to a symmetric $T\in CP_{M,N}.$
If $ r(T)$ is  affine  then
$r(T_{2}\circ T_{1})=r(T_{2})\circ r(T_{1}).$}\\
{\it Proof}\\
i) Requires a simple check.
ii) From previous Theorem  $r(T_{2}\circ T_{1})$ and $r(T_{2})\circ r(T_{1})$
are covariant maps of $CP({\cal S}_1)$. Let $\xi,\,\xi_{2},\,\xi_{1}$ be 
the associated scale factors 
One has to show that $\xi=\xi_{2}\,\xi_{1}$.
From equation (\ref{factor}) one finds indeed 
\begin{eqnarray}
\xi &=& \sum_{k=1}^N  M(T_2\circ T_1)_{ {\bf{j}}_1,\,{\bf{i}}_k}
= \sum_{l=1}^M\sum_{k=1}^N  M(T_2)_{ {\bf{j}}_1,\,{\bf{i}}_l}\,
M(T_{1})_{ {\bf{i}}_l,\,{\bf{i}}_k}\nonumber \\
&=&(\sum_{l=1}^M  M(T_{2})_{ {\bf{j}}_1,\,{\bf{i}}_l})\,
(\sum_{k=1}^N  M(T_{1})_{ {\bf{i}}_1,\,{\bf{i}}_k})= \xi_{2}\,\xi_1 , 
\end{eqnarray}
where we used the independence of $M(T_{2})_{ {\bf{i}}_l,\,{\bf{i}}_k}$
on $l.$
$\hfill\Box$\\
We conclude the section by 
a simple explicit computation, that shows the power of the notion of covariance.
Let us consider the case $d=2,\,N=1,\,M=2,$
with initial state
$\rho=2^{-1}({\bf{I}}+\sum_{\alpha=x,y,z}\lambda_{\alpha}\sigma_\alpha)$
(the $\sigma$'s are the Pauli matrices).
If $T\in CP_{2,1}$ is covariant and symmetric one must have
 $T({\bf{I}})\in{\cal A}_{2,2}=\mbox{span}\{ {\bf{I}}_2,\,C_2\}$
where 
\begin{equation}
C_2\equiv \sum_{\alpha=x,y,z}\sigma_\alpha\otimes \sigma_\alpha=
2\, P-{\bf{I}}
\end{equation}
is a traceless combination of the identity and the transposition
$P\,|\psi\rangle\otimes|\phi\rangle=|\phi\rangle\otimes|\psi\rangle.$
Moreover the $T(\sigma_\alpha)$'s must be totally symmetric operators
that transform  according the adjoint ($j=1$) representation of
$SU(2).$ The totally symmetric sector of $\mbox{End}({\bf{C}}^{4})$
is ten dimensional It is spanned by the elements of ${\cal A}_{2,2},$
($j=0$)
five operators  realizing a $j=2$ multiplet of $SU(2),$
and by the  $S_\alpha=2\,\Delta_2(
\sigma_\alpha), \,(\alpha=x,y,z)$ corresponding to $j=1.$
Therefore, from Theorem 2 [notice that trivially 
${\cal H}_{sym}^1={\cal H}$]
one has, $T(\sigma_\alpha)=\xi\, S_\alpha.$
Putting all together 
%\begin{equation}
$
T(\rho)=4^{-1}({\bf{I}}+ t\,C_2 +\xi
\sum_{\alpha} \lambda_{\alpha}\,
S_\alpha),
$
%\end{equation}
one has 
$$
\mbox{spec}\,T(\rho)=\{\frac{1}{4}(1\pm 2\,\xi+t),\,\frac{1}{4}(1-3\,t)\}
$$
form which, by imposing the positivity {\it and}
 optimality, one immediately gets 
(by {\it covariance alone}) the Bu\v{z}ek-Hillery result 
$t=1/3$ and $\xi_{max}=2/3$ \cite{BUHI}.
Notice that the optimal cloner has support in ${\cal H}_{sym}^2.$
\section{Summary}%%%%%%%%%%%%%%%%%%%%%%%%%
In this note it has been rigorously shown that the optimal (with respect to a  
metric criterion)  $N\mapsto M$ {\it pure state cloner} of a general $d$-dimensional 
quantum system can be described by a 
simple state-independent shrinking of the generalized Bloch vectors
associated to the reduced density matrices.
The structure of the proof can be summarized as follows.
Over the space $CP_{M,N}$ of $N\mapsto M$ cloners  a 'merit' functional is introduced 
in terms of the induced (non linear) maps of reduced (one-system) states.
This functional --which has a clear geometrical meaning in the setting of the generalized
Bloch representation (GBR) -- is concave and invariant under  the natural actions 
of the groups
$S_M,\,SU(d).$ 
This allows us to restrict our attention to covariant 
(i.e. invariant respect to the group action) cloners:
given a group orbit, by 'averaging' and using concavity,
one can build a covariant cloner with no worse quality. 
This cloner results to be universal (cloning quality independent on the input state), 
and the components of the associated GBR map satisfy an automorphicity contraint.
Allowing only for pure inputs and resorting  to the intimate connection
between representation theory of unitary and symmetric groups, 
one obtains the final result, that by linearity extends to the whole
space of states over the totally symmetric subspace of ${\cal H}^{\otimes\,N}.$
\begin{acknowledgements}
The author thanks D. Bru\ss, and  
 C. Macchiavello for introducing him to the
cloning problem, M. Rasetti for stimulating discussions and critical reading of the manuscript,
 Elsag-Bailey for financial support.
\end{acknowledgements}

%%%%%%%%%%%%%%%%%
\end{multicols}

\begin{references}
\bibitem{FID} 
Although this quantity has a direct physical meaning just for
pure states, it is well defined $\forall\rho\in{\cal S}.$
\end{references}
\end{document}